\documentclass[aip,twocolumn,letterpaper]{revtex4}

\usepackage{fullpage}

\usepackage{graphics}
\usepackage{epsfig}

\usepackage[margin=1.0in]{geometry}

\begin{document}
\title{Stellarators with enhanced tritium confinement and edge radiation control }
\author{Allen H Boozer}
\affiliation{Columbia University, New York, NY  10027}

\begin{abstract} 

A stellarator design is described with the purpose of achieving three goals:  (1) Enhance the confinement time of tritium. (2) Have a sufficient density of high-Z impurities to radiate the thermal power escaping from the core while having an extremely low impurity density in the core.  (3) Maintain a large fraction of the plasma in a burning plasma state with an optimal tritium fraction.  Some features of this design could be used in tokamaks.  Although having three confinement zones is natural for stellarators, it is not for tokamaks.   

\end{abstract}

\date{\today} 
\maketitle


\begin{figure}
\centerline{ \includegraphics[width=2.0 in]{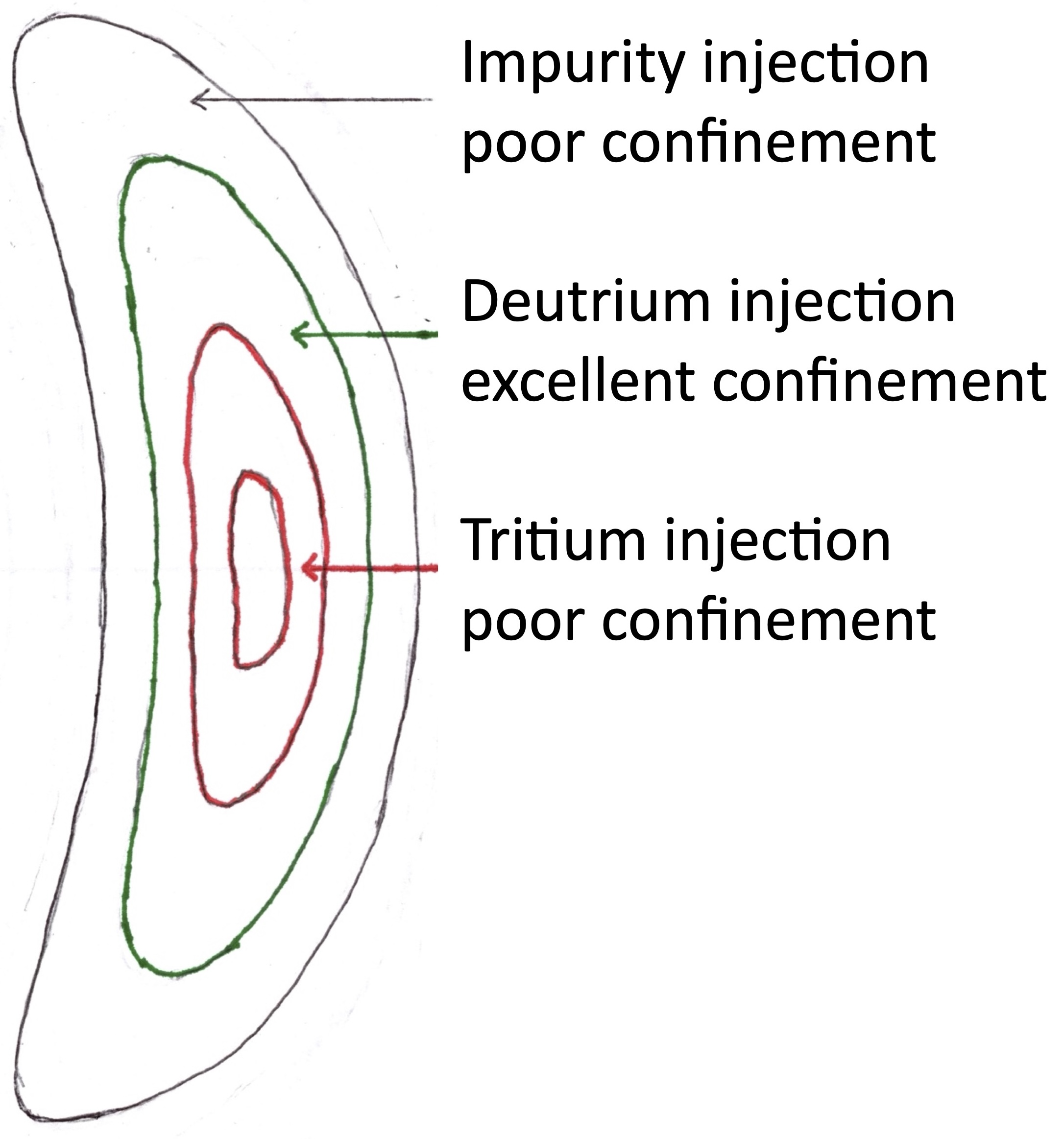}}
\caption{A stellarator plasma could be designed to have three confinement zones.  The innermost zone would have rapid transport and be the target for injected tritium pellets.  The intermediate zone would have sufficiently low transport for a fusion burn and deuterium pellets would be injected in the outer part of this zone.  The outermost zone would have rapid transport and extend into the region of open lines that form the divertor and be the target for injected impurities. } 
\label{fig:zoned}
\end{figure}

Three fundamental requirements on fusion power plants are that tritium be used efficiently, that the thermal exhaust be spread sufficiently uniformly to avoid damage to the walls, and that the helium produced by deuterium-tritium reactions be removed from the plasma chamber. 

Helium removal is generally envisioned to be accomplished by diverting the particle exhaust through relatively small apertures in the walls into divertor pumping chambers.   

The concentration of the particle outflow could produce unacceptable hot spots on the walls.  The heat can spread with acceptable uniformity over the walls by electromagnetic radiation \cite{Detachment 2017}.  But to achieve the required radiation, line radiation from high Z impurities is required.  The impurities required to radiate the power at the plasma edge cannot be allowed to penetrate to the plasma core because their radiation would be too great to be consistent with a fusion burn.

The fraction of a deuterium-tritium plasma that is tritium $f_t$ is generally chosen to maximize the fusion reactivity and minimize the required energy confinement time for a fusion burn.  When this is done, only a fraction of a percent of the tritium would burn during each cycle of the tritium through the plasma if the confinement time of the tritium equaled the energy confinement time.  In experiments the confinement time of particles is approximately an order of magnitude longer than that of energy \cite{Jakobs:2014}.  Nevertheless, the tritium losses during this recycling is a major issue in the tritium self-sufficiency of fusion power plants \cite{tritium,tritium2,{tritium3}}.  Ideally, the confinement time of the tritium would be infinite.  As discussed in the appendix to \cite{Boozer:2021}, the fraction of the tritium that is burnt in one cycle through a power plant can be made arbitrarily large by making the tritium fraction $f_t$ small, but at the price or requiring better energy confinement for a fusion burn.  The burn fraction in one cycle scales as $1/f_t$, while the required confinement for a sustained burn scales inversely with $f_t(1-f_t)$.  The required confinement is minimized at  $f_t=50\%$.  The burn fraction could be doubled at the cost of making the energy confinement 4/3 times better.

Tritium is extremely valuable.  A reliable source of tritium at a cost of less than ten million dollars a kilogram would make fusion economic  without a tritium-breeding blanket.  A plant that produces more tritium than it requires would find a ready market.  If excess production were possible, some DT power plants could be built without a breeding blanket, which would reduce the cost and size of those plants.   Deuterium can be reliably obtained at a price approximately three orders of magnitude cheaper.

A concept for addressing these fundamental power-plant issues is illustrated in Figure \ref{fig:zoned}.  This figure is highly schematic, and careful design would be needed to determine the optimal widths of the three confinement zones and at what depths the deuterium and the impurities are optimally injected.  One of the conditions on the central poor-confinement region is that it be reachable by practical pellets.  The poor confinement in the burning plasma core makes it easy to maintain an optimal tritium fraction $f_t$ throughout that region and ensures helium and other impurities cannot accumulate there \cite{Boozer:stell-imp}.   Ideally, the gradients of plasma temperature, density, and composition would be zero in the poor-confinement core.

In many models, keeping impurities out of the core requires an outward plasma flow to sweep the helium and other impurities from the plasma \cite{Jakobs:2014}.  This is far better done with deuterium, which has no supply issues and is orders of magnitude cheaper than tritium.  A sufficiently rapid outward plasma flow is required to exclude high-Z impurities from the plasma core even though impurities are needed at the plasma edge to spread the thermal exhaust broadly over the walls \cite{Detachment 2017,Boozer:divertors}. 

Some of the features illustrated in Figure \ref{fig:zoned} can be achieved in tokamaks, but the three confinement zones are more natural to stellarators.  In 1991, Garren and Boozer showed that quasi-symmetric stellarators have sufficient free parameters to achieve exact quasi-symmetry on one surface \cite{Garren-Boozer:1991} with increasing departures with distance away from that surface.  The standard stellarator transport codes seek optimal transport on a number of surfaces.  A stellarator with excellent confinement properties in an annulus with poor confinement properties on either side could be sought and may be easier to achieve than a stellarator with excellent confinement throughout its core. The outermost region includes in whole or in part magnetic field lines that do not lie on magnetic surfaces but lead to divertors.

The present state of theoretical and empirical knowledge seems inadequate for assessing much less optimizing the concept of Figure \ref{fig:zoned}.  Nevertheless, experiments in LHD \cite{Ida:2023} have shown deep hydrogen injection together with deuterium fueling by recycling can---but not always---produce essentially identical hydrogen and deuterium densities in the deep interior but with a deuterium density that is peaked near the plasma edge with the hydrogen density lower at the deuterium peak than in the deep interior.  The theory \cite{Ida:2023,Angioni:2021} is complicated by transport being microturbulent rather than neoclassical.  The focus of theoretical studies of transport has been thermal energy, but understanding the transport mixing of species is more important to the success of fusion at the present stage of understanding.

Present knowledge of the physics of pellet injection \cite{Pellets: 2018} also limits assessment and optimization.  In particular, the lack of localization of particle deposition complicates studies.  This lack of localization has two causes (1) the distance through which a pellet moves while it ablates and (2) the cross magnetic-surface spread of the material by drifts after if has ablated but before it has spread over the magnetic surfaces.

It is far easier to understand how a stellarator could have a poor confinement zone in its center and sufficiently high confinement zone to achieve fusion in an annulus than a tokamak.  Nevertheless, particle injection at various depths---even without distinct confinement zones---could increase the tritium confinement time in a tokamak.  Important experiments could be done in both tokamaks and stellarators with the results having implications for both.  The importance of a better understanding of how the presence of one species affects the profile and confinement time of another is given emphasis by consideration of the concept of Figure \ref{fig:zoned}.

\section*{Acknowledgements}

This material is based upon work supported by the grant 601958 within the Simons Foundation collaboration \emph{Hidden Symmetries and Fusion Energy} and by the U.S. Department of Energy, Office of Science, Office of Fusion Energy Sciences under Awards DE-FG02-95ER54333 and DE-FG02-03ER54696. \\








\end{document}